# Shared Secret Places: Social Media and Affordances

**Ella Hafermalz**
Business Information Systems
The University of Sydney
Sydney, NSW
Email: ella.hafermalz@sydney.edu.au

**Dirk S. Hovorka**
Business Information Systems
The University of Sydney
Sydney, NSW
Email: dirk.hovorka@sydney.edu.au

**Kai Riemer**
Business Information Systems
The University of Sydney
Sydney, NSW
Email: kai.riemer@sydney.edu.au

## Abstract

The Social Media application Strava is used by exercisers to track running and cycling activities. Strava is carried with the exerciser and displays trophies and leaderboards to reward competitive performance. We were prompted by an auto-ethnographic account of Strava use to examine the way in which a particular stretch of running track around a lake showed up differently to the runner once Strava was integrated into their running practice. We look to Gibson's relational notions of "affordances" and "niches" to understand this change in direct perception. We propose that these concepts have potential in helping us to research and understand the ways in which groups of Social Media users share and construct a similar experience of place in a way that is largely invisible to non-users. We consider some of the preliminary implications of this differentiated use of place and demonstrate the way in which a relational view of affordances helps us to make sense of this phenomenon.

**Keywords**
Affordances, Social Media, Mobile Technology, Place, Direct Perception, Ecological Approach

> **Prologue**
>
> *I ran with my phone in hand, headphones in ears. The lake came into view as the path wound on. Dodging oncoming pedestrians, my pace quickened. At the conclusion of my run, slowing to a stop, I quickly unlocked my phone, pressed 'finish', and immediately tapped the trophy – where had I placed on the Sydney Park Lake leaderboard?*

## 1 Introduction

This account presents the mystery (Alvesson et al. 2008) that inspired this research paper resulting from a researcher's experience of running with the exercise tracking application *Strava*. The mystery is captured as the *story of the lake*. While running around a lake with Strava, the lake was disclosed (Spinosa et al. 1999) to the runner as a *segment* that afforded competitive interaction with other users. This experience led to an investigative discourse with colleagues regarding how Social Media use came into play in how the researcher directly perceived the lake, which was now disclosed as a new place to move through. We approach this phenomenon by applying and extending Gibson's (1979) theory of affordances. We first show how Gibson's original conceptualisation is valuable in understanding how we use mobile technology. We also identify how the theory of affordances can be extended to allow us to better grasp the ways in which our experiences of place are changing as our networked technologies are carried with us. Theorising from an auto-ethnographic account, we propose that Social Media practices have implications for how groups perceive the environment and for how these socially connected groups act in and utilize the environment.

An example of this phenomenon is illustrated in one of the authors' use of the exercise application Strava. Through this example we show that our use of mobile Social Media involves a change in the way in which our environment is *disclosed* (Spinosa et al. 1999) as a set of *action possibilities*, in accordance with Gibson's 1979 ecological conceptualisation of affordances. Rather than placing the affordance *in* the technology we use affordances in the original relational sense as Gibson intended, to discuss how we directly perceive our environment as we move through it. We extend this original conceptualisation by comparing Gibson's notion of affordances with the way in which groups of users negotiate place over time through shared Social Media applications. We put forward Gibson's (1979) relational understanding of a "niche" as a "set of affordances" to theorise how groups of users *utilize* a place in certain ways that are shared. A place of private running becomes a shared place for competition. We suggest that this shared direct perception of a place is specific to certain Social Media user groups and that this shared understanding can be hidden from other groups who occupy a different niche in the same place.

## 2    Direct Perception: Gibson's Affordances

The term *affordance* was invented by Gibson in 1979 and refers to *how we directly perceive* our environment in terms *of action possibilities, as we move through it*. This emphasis on *direct perception* and *mobility* was a major contribution to our understanding of the way in which animals move through and make use of their everyday environment. Gibson (1979) argued that in order to understand how the (human) animal acts in its environment, we need to look beyond static theories of vision and perception that had been derived from laboratory experiments which used "the headrest, the bite-board…the darkroom', and as a result "depend on the subject's being willing to hold his eye fixed like a camera" (Gibson 1979). Gibson was instead interested in *ambulatory vision*.

The challenge was to develop a theory of how we see the environment around us "with reference to a *moving* point of observation" (1979). This emphasis on perception during motion was very important to Gibson, because it recognized that animals are always moving unless they are sleeping, or dead (Gibson 1979). In his affordance theory the animal's attention is dynamic rather than static because "when no constraints are put on the visual system, we look around, walk up to something interesting and move around it so as to see it from all sides, and go from one vista to another" (Gibson 1979). This understanding of perception in motion was one of Gibson's main contributions. Another was the insight that affordances are relational.

Gibson (1979) introduces his term in the following way: "the noun *affordance* is not [in the dictionary]. I have made it up. I mean by it something that refers to both the environment and the animal in a way that no existing term does. It implies the complementarity of the animal and the environment." The way in which the animal and the environment complement one another manifests in action: "the *affordances* of the environment are what it *offers* the animal, what it *provides* or *furnishes*" (Gibson, 1979, emphasis original). He offers a seat as an example. "If a surface of support…is also knee-high above the ground, it affords sitting on…it should *look* sit-on-able…if the surface properties are seen relative to the body surfaces, the self, they constitute a seat and have meaning" (Gibson 1979). The use of the knee-high scale implies a certain kind of animal – a human – and how it directly perceives its surrounding environment in terms of what it offers the human form and social actions.

Here Gibson carefully rejects the ontological subject-object dualism, clarifying that "an affordance is neither an objective property nor a subjective property; or it is both if you like. An affordance cuts across the dichotomy of subjective-objective and helps us understand its inadequacy… it is equally a fact of the environment and a fact of behaviour" (1979). Gibson here is setting up the notion of *direct perception*. In emphasising the unmediated, direct way in which we perceive the world, Gibson wishes to sidestep the cognitive mode of interacting with our surroundings to which the mentalism movement subscribes. In other words, rather than perceiving the environment as a collection of objects (e.g. chairs) that need to be translated in terms of meaning we will see action possibilities directly and unmediated (e.g. the opportunity to sit). Affordances, in Gibson's conceptualisation, are not reducible to subjective experience, neither are they locked in what is being perceived. So, "the behaviour of observers depends on their perception of the environment" but this does not imply that their behaviour "depends on a so-called private or subjective or conscious environment" (Gibson, 1979).

Gibson (1979) makes it clear that affordances depend on us, but are independent from any one *particular* subject. Affordances are always social, they do not exist because they are inscribed in the object, nor because of an observing subject, but because they are already socially shared in a way that has meaning for that particular animal. Affordances thus exist in the environment, yet not in the way objects exist. As such, his notion of environment refers to the socially interpreted environment, which carries the meaning regarding what objects afford and thus what they are.

How we experience our environment then is, according to Gibson, caught up in how we *act*. How we act is related to what we perceive, and what we perceive are affordances – that is, we see distinctions in surfaces, medium, and substances (Gibson, 1979) according to their *meaning*. These distinctions are disclosed according to what they allow us to *do*: what they *afford*. Affordances are intertwined with the physical environment but at the same time depend on human action. It is in the complementarity of the two that affordances reside. Affordances are thus *relational and performative,* embedded not in a dualistic relationship between a subject and an object, but enacted in the environment through practice. We return to Gibson's theorising on relevant aspects of affordances theory later, in response to the following auto-ethnographic account.

## 2.1　The Mystery of the Lake: From Running to Competing

At this point we return to our original mystery (Alvesson et al. 2008). The runner in our vignette found her way to a lake in a park, circumscribed by a path that people use for walking and running. As she began to use the exercise application Strava to track the speed and distance of her runs, she was awarded a trophy for achieving a competitive running time in a *segment* identified as "Sydney Park Lake". A leaderboard was displayed in the app which showed her time in comparison with other Strava users who ran the same segment. Over the next few runs, the lake was disclosed differently. The path no longer only afforded running, it now afforded competition. She was moving faster through this place, aware of others who had run here before. The fusion of the technology, a networked application user group and a running practice gradually became implicated in a shift in her direct perception of the path and her experience of the lake as place. How can we account for this shift, and how do we understand the role of mobile technology and Social Media here in the context of Gibson's theory of affordances?

Through the use of mobile technology and Social Media, an alternate lake was disclosed, one that was shared in *secret* with other users of the Strava platform. We present our mystery as follows: 1) how might the introduction of mobile technologies be implicated in a *change* in our direct perception while moving through place? 2) What implications does such change have for our experience, our behaviour, and the construction of place? We consider the Strava story here as an opportunity to make use of an extension of Gibson's relational theory of affordances – that is, how particular groups occupy, through their actions, specific *niches*. We suggest that this theorising can help us understand how groups of Social Media users are beginning to exhibit differential behaviour when moving through public places. Our theorising makes use of an auto-ethnographic account through which we demonstrate the nature and importance of this mystery. We conclude with guiding questions for future research.

# 3　Sharing Secret Places: Mobility and Social Media

We draw attention here to an emerging dynamic of how networked groups of people who are engaged in shared Social Media practices move through place in differentiated ways. The increased mobility and connectivity of computing devices has intersected with a rise in the use of Social Media. Social Media are defined as a "new class of information technologies" that "support interpersonal communication and collaboration using Internet-based platforms" (Kane et al. 2014). Based on both ideology and technical and interactional characteristics of Web 2.0 (Kaplan and Haenlein 2010), Social Media emphasise interactivity and the sharing of user generated content. Combined with wireless connectivity infrastructures Social Media make it possible for people to contribute, share, and interact with Social Media content while moving through the environment.

## 3.1　Moving through place

Exercise applications such as Strava and Nike+ are not frequently discussed in IS Social Media scholarship. However these GPS-enabled exercise tracking platforms represent growing networked communities that comes together to share, comment, and *like* each other's content in the context of fitness, exercise and competition. Consequently, an enormous data set is being created that makes these activity groups' behaviours visible, for example, to urban planners. These platforms are also interesting in the way in which they relate to our experience of moving through place.

For Gibson (1979) *place* is a "location in the environment". Places can be nested in other places. They can be named but "may not be demarcated with sharp boundaries" (Gibson, 1979). Places are recognisable and so the notion of *place* is a social one - its meaning is constructed through shared experience and social understanding of what that place affords. Animals of a similar *kind* experience a place in a shared way by moving through "the same paths" of its habitat over time (Gibson, 1979).

Because "the environment surrounds all observers in the same way that it surrounds a single observer," (Gibson, 1979) animals of the same kind perceive affordances similarly and so share the experience of place as a collective. There is an important connection here between the kind of animal, how it moves through the environment, and how it perceives and recognises place according to the actions that place affords.

### 3.2 Mobile connections

Bunce et al. (2012) explain that although Social Media platforms have been found to mainly develop existing "offline" relationships, there are times when events (such as natural disasters) construct opportunities for "an increase in new relationships between (otherwise unknown) producers and consumers of information." We suggest that such relationships are also developing around everyday events and activities in a way that is tied to how people use and experience particular places. The exercise application *Strava* for example connects "otherwise unknown" (Bunce et al. 2012) users around a shared experience of a particular segment of shared path.

Cyclists who ride the same paths as other Strava users have their rides linked in the app, so their chosen avatars and their performance statistics become shared and visible. Although activity groups have always discussed experiences and shared tips and achievements, apps such as Strava layer this experience over graphic representations of place with increased immediacy, detail, visibility and interaction possibilities. It is possible to immediately compare one's own performance with both known and unknown others, and the process of disclosing the "significant places" (Gibson, 1979) known as segments is silent – it can even occur unwittingly, as will be shown.

## 4　Running 'Together' with Strava

> *I had unintentionally discovered that the lake was not only a part of my usual route, but a particular part of this route that was involved in a competition. I began to act differently at this point of my run. As the lake is at roughly the halfway point, I am usually tired by the time I reach it. To avoid this fatigue I found myself slowing down in the relatively unmonitored sections of my run prior to the lake segment. This allowed me to run more quickly during the part of the run that counted for competitive status against others who were using the Strava application in this place. In a sense, those other users were with me in the race, it was as if I joined them when I reached the lake.*

### 4.1　Introducing Strava

The auto-ethnographic account that follows is based on an 18-month period of use of the iPhone version of the mobile application *Strava*. Strava advertises itself with the tagline "Fast, far and free — with Strava, you're never alone" (Strava 2015). When activated, the free version of Strava tracks users' running or cycling activities and records measurements such as distance, average pace, and location. These statistics are then compiled in a dashboard that presents a map of the completed activity as well as how your performance ranks against other users. Strava (2015) describe this functionality as follows: "Strava lets you track your rides and runs via your iPhone, Android or dedicated GPS device and helps you analyse and quantify your performance. Strava provides motivation and camaraderie, and helps us prove that we're out there doing what we love to do." The slogan's emphasis on "camaraderie" and demonstration of activity speaks to the social aspect of this platform. The application conforms to Social Media conventions by encouraging users to *follow* one another. The user is represented by a small round profile picture. Their activities are presented in snapshot format on a social feed and users can give each other *kudos* by clicking a *thumbs up* button that appears below each completed activity.

There are also elements of gamification (Deterding 2011) in the platform, such as the designation and sharing of segments. These segments represent an opportunity to compete with other users who also run or cycle on that stretch of path. Individual performance on this stretch is compared with other users and a leaderboard displays names and ranking. This occurs whether or not you sign up for the segment, and your status is visible to others. After each activity the user can see how their activity compares to their past performance and to the performance of others. If users have travelled on the same path at the same time, Strava links these activities and imports photographs that were posted to Instagram during the activity. In this way, Strava conforms to the Social Media tenets of content creation and social interaction (Kaplan and Haenlein 2010). Empirical material from Strava use is now analysed using theoretical counterpoints in the final section.

During the 18 months of being a runner with Strava the lead researcher constructed breakdowns in understanding (Alvesson and Kärreman 2007) that stemmed from a familiarity with Gibson's work on affordances, prior conceptions of the use of Social Media, and empirical material that challenged some of the assumptions that go with extant theoretical frameworks. Further reflection and investigation has been undertaken, resulting in, at this stage, the construction of a "mystery" and the analysis of "(a) the broader relevance of an empirical finding, (b) the problems with the earlier theory or critique, and (c) some hints of a new understanding through the formulation of the mystery" (Alvesson and Kärreman 2007). In this section we continue with an account of the experience of breakdown, and the researcher's subsequent mystery construction. In keeping with the approach outlined by Alvesson and Kärreman (2007), the personal account in this section does not take a grounded approach to theorising (Eisenhardt 1989), but rather interweaves both empirical material and theory while considering these as complementary in the research process known as abduction (Alvesson and Kärreman 2007; Timmermans and Tavory 2012).

## 5   An Auto-ethnographic Account of the Lake

Being a regular but by no means competitive runner I was curious about Strava for two reasons – firstly, I wanted to experience the tracking aspects of the application first hand and secondly, I wanted to see whether these mechanisms and capacity to see progress would motivate me to run more regularly. I have now been using the application for 18 months and it became an integral part of my running routine. The following observations summarise my experiences of using Strava in line with the themes that I identified from my observations and notes.

### 5.1   Becoming acquainted

In the run shown in Figure 1, the red line represents the path taken. This does not join up because I started and finished using the Strava application some distance from my house. Given the regularity of my runs, I initially did not want publicly available information to reveal my home location through the mapping feature. To further protect my privacy, I at first rejected the Social Media aspect of the application, by using a pseudonym and not following other users or allowing users to follow me. Thus the application was not a social tool, rather it was a tracking and record keeping device with a competitive game element that was inward-facing. This was is in keeping with McGonigal's (2011) view that "sociability and community are optional features of the game structure and are only peripheral enhancers of the core game mechanics" (Dery et al. 2014). The aim was to make running more gameful (Deterding et al.), not to compete with others.

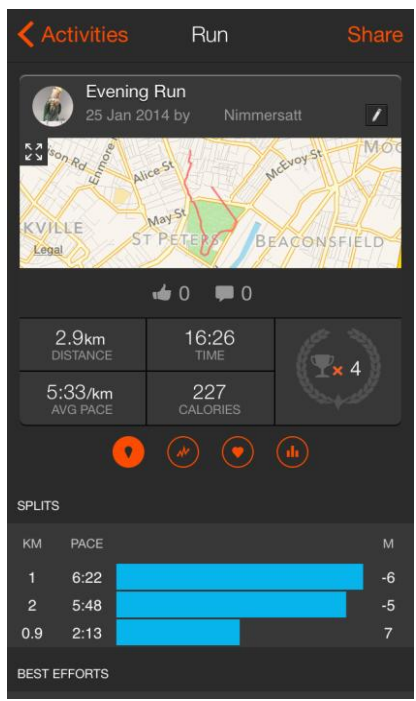  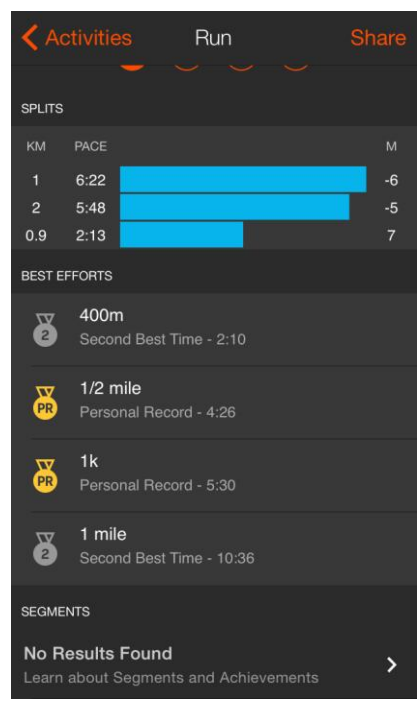

*Figure 1: Summary of Run*　　　　　　　*Figure 2: Summary of Performance*

A trophy was displayed on screen after one of my first runs (Figure 1). By investigating further (by scrolling down – Figure 2.) I could see I had beaten my previous times over certain standard distances, a record that was detached from any particular place. This self-competition was at first motivating, until I stopped beating my previous records and no longer received the trophy icon. As I ran less, I knew I would be less competitive against my past self and I lost interest. Some weeks later however my curiosity was renewed through discussion with friends who used Strava for cycling. I already carried my phone with me to listen to music, so the effort of including Strava was minimal and I activated it once again.

## 5.2 Encountering 'unknown others'

Even though I had rejected the notion of social interaction on Strava, I was brought into this networked competition through the device of segments - a central component of the Strava eco-system. After exploring the screen to investigate what trophies had been awarded after a run, I noticed that I had been added to a leaderboard that featured the names and times of other female runners under the heading "Sydney Park Lake" (Figure 3). I had, without knowing it, run the fifth fastest time around a lake that had been in my route for some time. This revelation was at first confusing - I had not intended to make my activities public and felt irritated at being made visible. Curiosity took over however and I began to focus on the game.

The next time I ran, my activity was more focused on the lake as segment. It became an opportunity to compete and I sped up accordingly. Although I ran alone, unknown others were running with and against me. Paharia (2013) relates this game element of mobile applications to the quantified-self movement. This movement is seen to be intersecting with the capabilities of Social Media to create an opportunity to engage people in particular activities to "collaborate and compete with their networks to better achieve individual and collective goals" (Dery et al. 2014).

## 5.3 The lake as segment: a shared secret place

Sydney Park Lake had at some stage been submitted to Strava as a segment which then stands in for a competitive stretch of path. This process is described further in the screenshot (Figure 4) taken from the help screen for 'What's a segment?' on the Strava iPhone application. Through the process of nominating or running segments, a section of a road or trail becomes specific and available to those who are enrolled in the Strava segment practice as affording competition. Even while rejecting the social interaction that Strava promotes, I had entered into a competitive relationship with unknown others and "Sydney Park Lake" became differentiated in my run (Figure 3). The path around the lake was now a segment, and was linked to the practice of competition, status and rank as displayed in the leaderboard screenshot below (Figure 5 – first names and images are blurred to protect privacy).

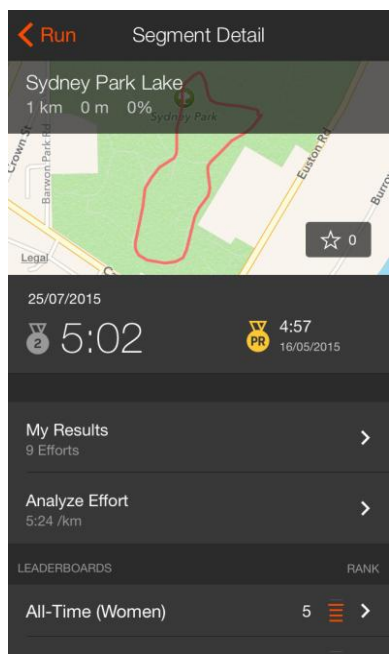 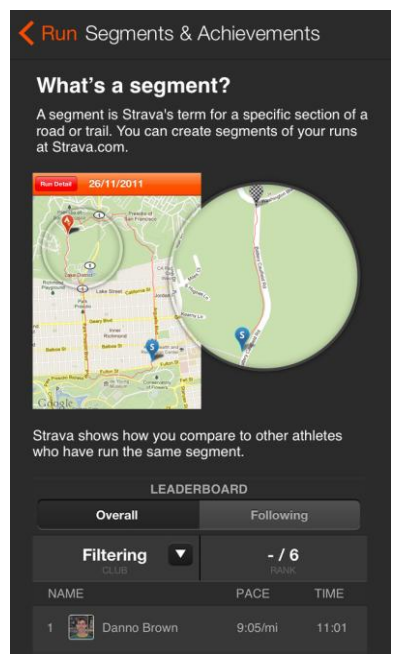 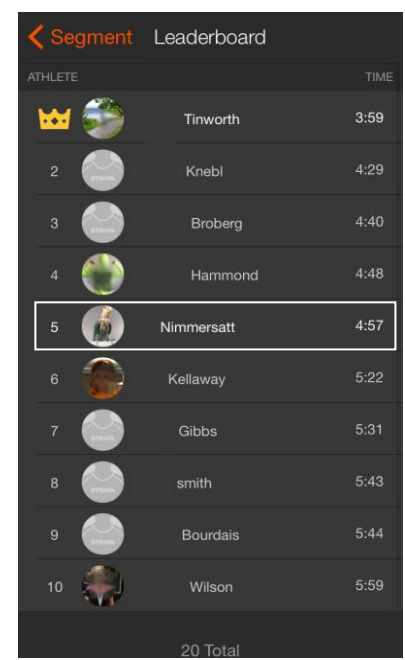

*Figure 3: The Lake*　　　　　　　　*Figure 4: Segments*　　　　　　　　*Figure 5: Leaderboard*

What I started to observe was that my use of Strava over time *changed how I perceived the lake as place*, in that it became differentiated in terms of *affording competition*. This is not to say that my experience of the lake was mediated by my use of the application, rather my behaviour changed as did my movement through the place as I began to directly perceive the lake and surrounding path differently. This would suggest that rather than the gamification characteristics being something that existed *inside* the application, the game structure was instead involved in a relational understanding of what action opportunities a particular place afforded.

This understanding is aligned with Gibson's (1979) theory of affordances, and yet it is not entirely explained by it. This is the stage at which a mystery began to emerge – how does our direct perception of place metamorphose as Social Media becomes integrated in our activities? The following section begins to join the "micro to the meta" (Riordan 2014) and offers some preliminary analysis in response to this mystery.

## 6  Discussion

The above example illustrates how the relationship between the researcher and the lake metamorphosed – where once the lake was a part of the exercise routine it was now disclosed as affording competition. The change in affordance here is related to the researcher's direct perception of the lake shifting in some way alongside the mobile technology that she held in her hand as she moved through the lake-as-place. To explain this change we extend Gibson's theory of affordances into a broader argument by introducing the concept of ecological niches, and demonstrate how this concept may help us understand how different user groups share certain experiences and utilizations of place.

### 6.1  Tools as extensions of the self

Gibson (1979) argued that perception of the everyday environment is direct rather than mediated by representation, in that our familiar everyday environment is always perceived in terms of action possibilities. This concept of direct perception in terms of affordances is a powerful and useful one for IS research and we wish to highlight its importance here. Where we reach a potential roadblock however is in what to do with the *change* in direct experience that is recalled in our researcher's auto-ethnographic account. If Strava is not *mediating* the direct experience of the runner, how can we interpret the change in perception? We turn to Gibson's description of tools and their relationship to the human animal to better grasp this phenomenon.

Gibson echoes philosophers such as Heidegger (1962) and Merleau-Ponty (1962) in his treatment of tools. Gibson (1979) writes that:

> When in use, a tool is a sort of extension of the hand, almost an attachment to it or a part of the user's own body, and thus is no longer a part of the environment of the user. But when not in use, the tool is simply a detached object of the environment, graspable and portable, to be sure, but nevertheless external to the observer. This capacity to attach something to the body suggests that the boundary between the animal and the environment is not fixed at the surface of the skin but can shift…When being worn, clothing, even more than a tool, is a part of the wearer's body instead of a part of the environment.

This understanding of tools-in-use becoming an appendage (Malpas 2012; Wilken and Goggin 2012) of the human body resists the notion that the phone is merely mediating the runners' experience. Rather, the runner is *fused* with the phone and the Strava platform in a way that is important for the runner's actions and perception as they move through the environment.

The phone in the runner's hand is in this view not an object carried by the subject, rather it effectively becomes a *part* of the runner's body. This fusion does not just refer to the shell of the phone but also to what is *in use* – in this case the Strava application, its GPS positioning and timer, its reference segments and Social Media connections. If we compare our runner with another person who is walking phoneless around the lake, to what extent would we expect them to perceive the lake in the same way? While this environment offers the walker a suitable surface for walking, the runner with Strava extends this and perceives the lake in terms of speed, competition, and in some way in relation to the line that is emerging on a digital map that will be shared with unknown others. As affordances refer to complementarity between the perceiver and the environment, we argue that the reconfiguration that occurs when technologies are fused with the perceiver may also involve an emergent disclosure of the environment and therefore the affordances that appear.

## 6.2 Social animals

The iPhone affords carrying and is similar to other portable artefacts which, when worn can be thought of as becoming a part of the body. We could postulate that the iPhone, carried in the hand, has become an extension of the runner's perceptive self. Even though there are no physical markers at the lake that signal a start-or-stop point, our runner is attuned to the environment differently because they are enrolled in a particular practice that involves the equipment of social competition. The consequence of this fusion is that in some sense, the runner is no longer what they were – they are now runner-with-Strava rather than the runner they were before. But importantly they are not even a solo runner as they run alongside invisible others.

This observation suggests that we move beyond the individual runner to consider the networked group of Social Media users who utilize the same environment in similar ways. Groups of Strava users who utilize the lake for competition are moving through this place differently to others who frequent the lake. Perhaps then we can even understand that the runner-with-Strava is a somewhat different *kind* of animal from the walkers she passes. However, according to Gibson (1979) there is only "one world" and also only one "human animal" who perceive affordances fairly homogeneously. While some minimal learning (Gibson, 1979) might be involved in adjusting our understanding of affordances as we engage in practices, our relatively similar ways of moving through the environment transpire in relatively similar perceptions of affordances.

How then do we account for our researcher disclosing the lake-as-competition when others might disclose the lake-as-walking path, while animals such as birds might disclose it as lake-as-nesting place? This comparison to other animals is perhaps the key here. For example a bird is socially enrolled in the practice of perching and is appropriately equipped for perching, so it would perceive and understand an appropriately shaped stick not as a *stick* but as a *to-perch* (Heidegger 1962), in a way that wouldn't show up to a human (without concerted effort). A similar narrative can be told about how people as human animals move through their environment with equipment. The practice of running for example requires running shoes. Without attaching these shoes to their feet, the gravelled path would likely not be disclosed as affording running.

This is not to say that the shoes *mediate* the runner's direct experience, rather that they are a part of what it is to *be* a runner. Many more examples could be offered to illustrate this diversity of fusions – consider for example the cyclist, who relies on their bicycle to be a cyclist, and as a (successful) cyclist, perceives the environment in terms of cycling. Surfaces appear differently in cycling motion than in running motion – the same environment is directly perceived in a different way. We do not need to constantly remind ourselves which equipment we are using – the road is perceived differently and is dealt with accordingly. This is an example of complementarity between animal and environment, though it is a different conception of animal to Gibson's. At first it seems that Gibson gives us little in the way of a framework for making sense of these shifts, of this fluidity of perception. Upon closer reading however we propose that the ecological example of "niches" as outlined by Gibson (1979) is a potentially valuable concept for understanding such shifts, where groups with different tools-in-use perceive and utilize places in diverse and evolving ways.

## 6.3 New niches disclosed

Ecologists recognise that animals occupy or utilize different parts of the environment and differentiate these places with the term *niche*. This is not the same, Gibson points out, as the species' habitat – rather "a niche refers more to *how* an animal lives than to *where* it lives" (Gibson 1979). Because of this emphasis on *how* an animal lives – its actions and doings, Gibson (1979) concludes that a "niche is a set of affordances". Consider again the lake – for the bird, it is a to-nest, to-swim, to-eat – this is the set of affordances that *together* construct the *niche* of actions that the bird inhabits. The walker instead sees the lake and its surroundings as a to-walk, to-relax, to-talk – this set of affordances is the walker's niche. We suggest that through a (not well understood) socialisation process involving Social Media, the runner-with-Strava comes to inhabit the same environment as others but *does* so differently and therefore occupies a *different niche* within that same environment. For our researcher, the lake and its surrounds is now disclosed as a to-compete and as a for-status. The runner-with-Strava is not the only kind of group to inhabit this place but her *niche*, which is shared with other Strava users, is distinctive.

Although we have only one natural environment, Gibson (1979) is adamant that it "offers many ways of life" and that "different animals have different ways of life". While Gibson is categorising different animals in a traditional sense, we could define this difference more broadly in order to see where his theory of niches can take us. If for example "the niche implies a kind of animal, and the animal implies a king of niche" (Gibson 1979) we can begin to grasp how direct perception of the environment and consequently how affordances are perceived are open to change through fusions that can be thought of as a new kind of animal. Depending on the configuration of the animal, its practices, and the way it moves through its environment, places are disclosed differently.

### 6.4 Consequences and implications

Visible markers of this differentiated way in which groups of technology users move through place are already emerging. At the University of Utah for example, "texting lanes" have been painted on footpaths for students who are frequently moving through places while looking at their phones (O'Neil 2015). These lanes have been created in response to frequent collisions and interruptions to pedestrian traffic and the desire to cater to the student demographic (O'Neil 2015). Similar lanes have been trialled in Antwerp, Washington DC and Chongqing (Aubrey 2015). The example of texting lanes illustrates how a group of people who are fused with their phones and messaging systems move through place differently, they *utilize* place in a particular way. These "textwalking" (Aubrey 2015) pedestrians, with their visual attention attuned differently to those who are not walking while looking at their phones, are inhabiting a niche that has been made visible to others with the texting-lane markers.

Another example from Social Media is the practice of sharing square-format photographs on Instagram. We propose that those who are enrolled in Instagram as a practice may move through places with an attunement to what affords Instagramming. It follows that vendors and sites such as museums, cafes and galleries that are deemed *Instagrammable* are likely to attract people who are enrolled in the practice of Instagramming. What we begin to highlight here is that groups form around Social Media practices in such a way that has implications for how they directly perceive and thus move through and utilize place. In this example, we wish to show that although someone may not be looking at their phone screen while walking, their direct perception of the environment and what it affords may be somewhat different to a fellow pedestrian who is not *concerned* with Instagramming. In this way, Instagrammers come together using Social Media in such a way that they disclose the environment differently and utilize a different set of affordances – i.e., a different niche, from non-users. This has implications for how the environment is shaped by and for that group and how they come to occupy similar places without direct coordination.

An intriguing aspect of this phenomenon of how Social Media, affordances and place come together is the way in which groups come to share a niche that is largely invisible to others who occupy the same place but utilize it differently. For example, consider watching from your car as a group of cyclists speed up on a stretch of flat road, then slow down as if in sync, all without saying a word. What just happened? The cyclists are enrolled in a practice of Strava use that discloses this segment of road in terms of competition. Without observable coordination, they have shared a direct perception of the road-as-competition in a way that is not clear to those outside of the practice. These cyclists-with-Strava have just occupied the same habitat as the driver-with-car but have perceived the affordances of the road differently – they have utilized a different set of affordances, a different *niche*. The niche is gradually disclosed through the shared practice of Social Media, while remaining mostly invisible to outsiders.

We can begin to consider the wider implications of this phenomenon. Apart from the way in which users gather around these shared niches, businesses behind applications like Strava also collate and sell the enormous amount of data generated by its users. It can be realistically anticipated that these data sets will over time be purchased and taken into account by developers and planners. Knowing where and when cyclists ride could potentially impact the planning of cycle lanes, which in turn may reinforce where riders ride. Popular segments and stretches could be attractive sites for complementary businesses. A conceptualisation of ecological niches for Social Media user groups could be helpful in analysing such phenomena. Of course more research is needed to investigate these propositions.

## 7 Conclusion

In this paper we were inspired by auto-ethnographic research to give particular attention to how fusions with mobile technology relate to a shift in the direct perception of the environment when in motion. Such a conceptualisation is very different from considering what a technology *contains* or how it *mediates* our experience. We show how this relational view of affordances and the associated concept of niches offers us a framework for considering the ways in which mobile technology and groups of people are becoming fusions, facilitated by Social Media, to create groups that perceive their environment differently and therefore act differently as they move through it. The concept of groups inhabiting niches as sets of affordances is particularly appealing here as a means of further investigating how these emergent groups disclose the environment in specific ways.

We offer here an account of a process that we are yet to fully grasp, in which the Strava user negotiates and becomes enrolled in the practice of Strava, and thereby eventually discloses (Spinosa et al. 1999) that group's "significant places" (Gibson 1979) ( i.e. which stretches of path are known and accepted as segments, and so afford competition, are anticipated and celebrated). We propose that this Strava group may simultaneously adjust how they move through that environment, to better and more fully utilize the niche or set of affordances that runners-with-Strava occupy. To others however, who are not "on" Strava, this behaviour could be seen as incongruent or inexplicable given the relative invisibility of the reconfiguration. The ways in which secret places become shared through Social Media and the implications of this shift in experience is a mystery that we put forward as a guiding phenomenon that deserves further attention and investigation.

We contribute a research orientation that examines the relationship between mobile technology, Social Media, and place. We suggest that Gibson's conceptualisation of affordances is critical in providing us with a language to discuss this dynamic. By explicating and adapting Gibson's concept of niches and viewing tools as extensions of the body, we provide a point of entry to theorise how Social Media users group together around activities to utilize sets of affordances in a way that, over time, will have implications for the places they inhabit. We anticipate that this issue will remain salient for the IS community in future research.

## Copyright